\documentclass[10pt,conference]{IEEEtran}

\usepackage{amsmath,amssymb,tabularx}
\usepackage{graphicx,cite}
\usepackage{epstopdf}
\usepackage{color}
\usepackage{url}
\usepackage{verbatim}
\usepackage{multirow}
\usepackage{gensymb}

\usepackage[draft,footnote,nomargin]{fixme}
\usepackage[footnote,nomargin]{fixme}

\begin{document}

\title{Traffic profiling for mobile video streaming}

\author{\IEEEauthorblockN{Dimitrios Tsilimantos, Theodoros Karagkioules, Amaya Nogales-G\'omez, and Stefan Valentin}
	\IEEEauthorblockA{
		Mathematical and Algorithmic Sciences Lab, France Research Center\\  
		Huawei Technologies France SASU\\ 
		\{dimitrios.tsilimantos, theodoros.karagkioules, amaya.nogales.gomez, stefan.valentin\}@huawei.com}}

\maketitle

\begin{abstract}
This paper describes a novel system that provides key parameters of HTTP Adaptive Streaming (HAS) sessions to the lower layers of the protocol stack. A non-intrusive traffic profiling solution is proposed that observes packet flows at the transmit queue of base stations, edge-routers, or gateways. By analyzing IP flows in real time, the presented scheme identifies different phases of an HAS session and estimates important application-layer parameters, such as play-back buffer state and video encoding rate. The introduced estimators only use IP-layer information, do not require standardization and work even with traffic that is encrypted via Transport Layer Security (TLS). Experimental results for a popular video streaming service clearly verify the high accuracy of the proposed solution. Traffic profiling, thus, provides a valuable alternative to cross-layer signaling and Deep Packet Inspection (DPI) in order to perform efficient network optimization for video streaming.
\end{abstract}


\section{Introduction}
\bstctlcite{IEEEbibChanges:BSTcontrol}
\label{sec:intro}
In 2016, video accounted for 60\% of the global data traffic in cellular networks -- a proportion that is expected to increase to 78\% by 2021 \cite{cisco2016}. Most of this traffic is generated by major video-on-demand (VoD) streaming services \cite{Sandvine} and based on HAS techniques, as standardized in Dynamic Adaptive Streaming over HTTP (DASH) \cite{DASH} and specified in HTTP Live Streaming (HLS) \cite{HLS}. Although the fast deployment of 4.5G is helping mobile network operators to support current streaming rates, major challenges such as 4K resolution, 360\degree{} live videos, and cloud rendering lie ahead.

There is now a wide consensus that stalls are the dominant Quality-of-Experience (QoE) factor for mobile HAS, as they quickly cause viewers to abandon their video session \cite{HAS_QoE}. This high impact of the video traffic's real-time requirement, as defined by the level of the play-back buffer, is often coupled with a greedy quality selection of HAS clients \cite{Sieber2015}. Once the video traffic is served with priority, the HAS client will increase the quality, thus, requesting a higher and higher rate from the network for fluent streaming. This combination makes it challenging to provide sufficient streaming quality while simultaneously satisfying other traffic types, such as voice and best-effort data, over the shared resources of a mobile network.

Interestingly, HAS traffic has fundamental characteristics, which can be exploited for more efficient scheduling \cite{6948362,ownICC16}. Schedulers with knowledge of video key parameters, such as video bit-rate and buffer level, can adjust their priority weights and improve the overall QoE in the network. Recently, rate throttling mechanisms have been deployed \cite{bingeonPaper} as an early attempt to actively manage video traffic in mobile networks. However, this approach suffers from the limits of DPI, which requires the close cooperation with service providers to either tag the video packet flows or to weaken end-to-end encryption. Even with such cooperation, DPI allows simple spoofing techniques by replacing tags via HTTP proxies and, thus, leads to an arms race between operators and malicious users \cite{bingeonPaper}. Consequently, network optimization and traffic management need a method to provide video parameters to the network layer in real time, without the need for tagging or for violating end-to-end encryption.

In this paper we address this problem by proposing \emph{traffic profiling}, which fundamentally differs from DPI and cross-layer signaling. Instead of extracting or receiving information from the higher protocol layers, traffic profiling only observes a packet flow from the downlink transmit queue of the serving base station (BS), edge-router or gateway. This observation is entirely done at the Layer 2 or 3 of the ISO/OSI protocol stack, with the aim of detecting characteristic features of video traffic. Based on these features, application-layer parameters of the ongoing video stream are extracted. We will demonstrate that this method estimates the current phase and bitrate of an HAS session with very high accuracy, only by observing the time stamps and the size of IP packets. 

Compared to DPI and cross-layer signaling, the benefits of this new approach are (i) no dependency on tags or cross-layer information that could be manipulated, (ii) no required standardization of cross-layer interfaces, (iii) no direct access to the application-layer, which may violate user privacy, and (iv) no need for weakening end-to-end encryption via TLS or Secure Sockets Layer (SSL). 

The main drawback of traffic profiling, however, is that it can provide only an estimation of application layer parameters, which may be erroneous. To this end, the focus of this paper is not only on the description of a traffic profiling solution, but also on its rigorous experimental validation. To this end, we conduct an extensive measurement campaign for the YouTube streaming service, which is a dominating source of global HAS traffic \cite{Sandvine}. Note that this study comes at no loss in generality, as our solution works for any kind of HAS traffic, such as DASH and HLS, with or without encryption.	After studying more than 170 hours of TLS-encrypted HAS traffic, we found a very high accuracy for the proposed solution, which opens a wide range of practical applications such as network optimization, traffic shaping and QoE-estimation.

According to our knowledge, this is the first work that proposes a traffic profiling solution for video streaming. Nevertheless, several interesting studies on the characteristics of HAS traffic were published. The authors in \cite{Ameigeiras2012} formulate a traffic generation model for a YouTube server, based on measurements from a computer and re-evaluate the model for mobile devices in \cite{CharMobileYouTube}. In \cite{Rao2011}, a model for the aggregated traffic of various streaming strategies used by YouTube and Netflix is proposed. As QoE models are becoming very valuable for network performance validation, Wasmer et al. provide such a study in \cite{Wamser2016} based on subjective tests for YouTube. Their study introduces a network traffic model for the YouTube control mechanism and analyzes YouTube's operation from an end-user QoE perspective. The QoE of YouTube traffic is also studied in \cite{Liu2015} using subjective tests and with an experimental setup similar to ours. Focusing on the redundant traffic and greediness of YouTube, Sieber et al. provide a very interesting study in \cite{Sieber2015}.

The remainder of the paper is organized as follows. We describe the main design of the traffic profiling solution in Section \ref{sec:model} and detail the algorithm in Section \ref{sec:VRSE}. In Section \ref{sec:results}, we present our experimental results and conclude the paper in Section \ref{sec:conclusions}.

\section{System Design}
\label{sec:model}
The proposed system simply observes the packet flow of a HAS session at the edge of a mobile network. Note that our traffic profiling solution has to operate at the edge, since (i) the estimators are based on the assumption that the observed packet inter-arrival times (IATs) are close to the end-to-end IATs and (ii) we exploit the fact that user-specific queues are available at typical BSs. Once these two criteria are met, the proposed solution can be deployed at any edge-router, gateway, BS or access point for various applications.

\begin{figure}[t]
	\centering
	\includegraphics[width=1\linewidth,trim={0.5cm 0cm 11cm 8.5cm},clip]{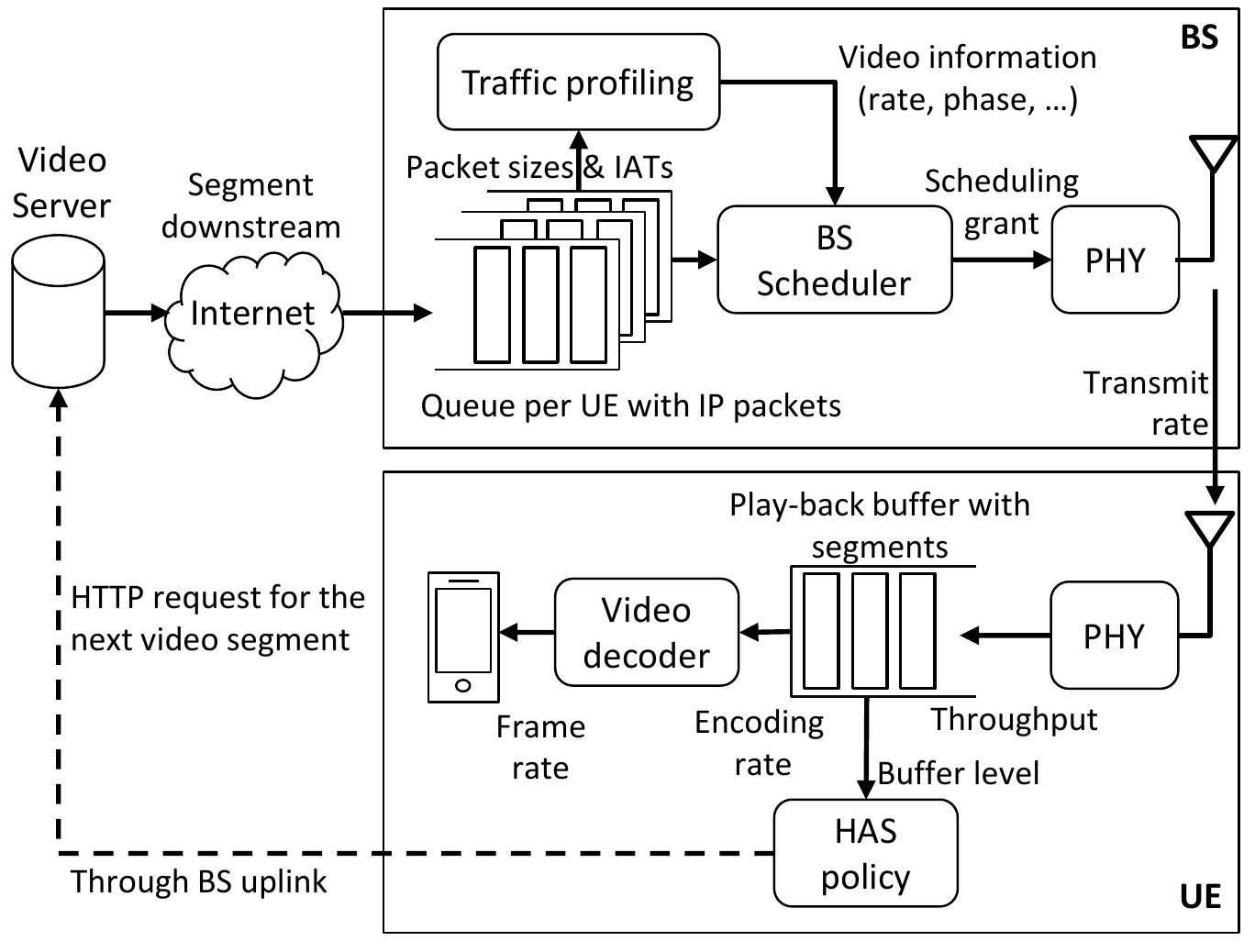} 
	\caption{Application example: Video-aware scheduler with traffic profiling for adaptive streaming.}
	\label{fig:System}  
\end{figure}

\figurename{ \ref{fig:System}} shows such an application example. Here, traffic profiling runs at the BS in order to provide its scheduler with video-specific information. While the design of a video-aware scheduler goes beyond the scope of this paper, a large body of research shows that more accurate cross-layer information allows to allocate radio resources more efficiently to the current traffic demands \cite{6948362,ownICC16}.

As in any video streaming system, the client application at the user equipment (UE) employs a play-back buffer to compensate for short throughput reductions and instantaneous increases in video encoding rate. In HAS systems, a single video sequence is subdivided into segments of a constant duration of several seconds. Each segment is stored at the video server in multiple quality levels and requested sequentially by the client. For each segment, the client adapts the quality and the request time depending on the transmission control protocol (TCP) throughput and the play-back buffer level. At the BS, arriving IP packets are placed in user-specific queues, from which packet size and IATs are observed and used for traffic profiling.

\begin{figure}[t]
	\centering
	\includegraphics[width=1\linewidth,trim={0.8cm 0cm 1.2cm 0.5cm},clip]{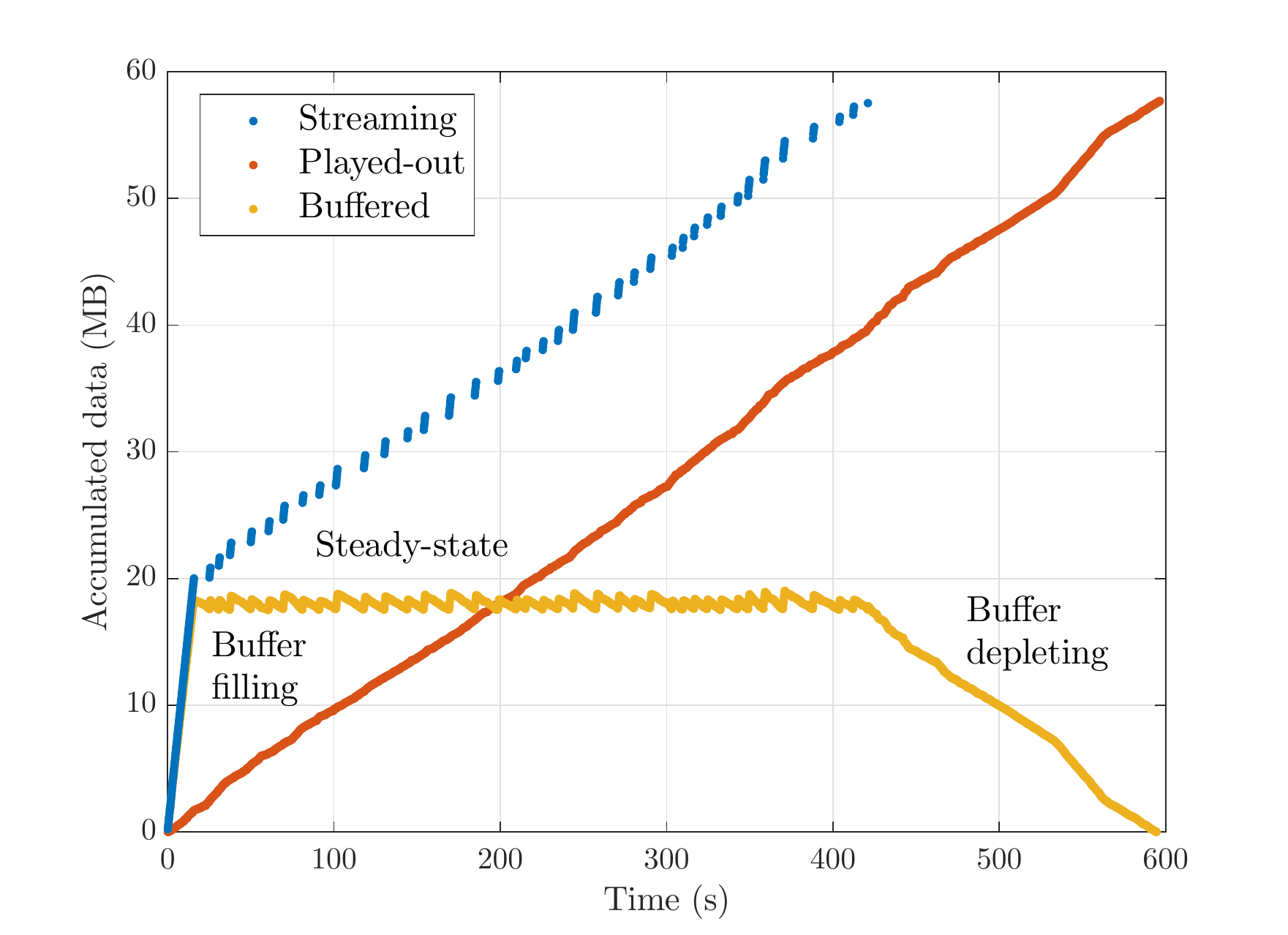} 
	\caption{Example of streaming the Big Buck Bunny movie \cite{BigBuckBunnieMovie} with 480p quality, using the setup in Section \ref{sec:setup}.}
	\label{fig:StreamingExample}  
\end{figure}

The downstream of a typical HAS session is shown in \figurename{ \ref{fig:StreamingExample}}, as measured according to Section \ref{sec:setup}. This figure shows the accumulated data over time for (i) TCP payload of streaming data, as observed in the BS downlink transmit queue and (ii) played-out data, acquired by analyzing the downloaded video segments. The subtraction of these two plots then provides (iii) the play-back buffer level. 
 
From this example we can easily distinguish the three main phases of an adaptive streaming session by looking at the slope of the plots. Firstly, we observe that there is an initial burst of data where the streaming rate is much higher than the played-out rate. This time period is denoted as the \emph{filling phase}, where the HAS client requests maximum throughput to quickly fill the buffer to a certain level. Once this level is reached, the client changes to the \emph{steady-state phase} in order to match its requested streaming rate to the video encoding rate. HAS achieves this rate match by a characteristic on-off request pattern, which leads to short packet bursts, followed by periods without packet transmission. These burst-wise requests control the streaming rate and keep the buffer almost steady around a target level. Finally, after the entire
video is transmitted, the session ends with the \emph{depletion phase} by playing-out the remaining bits from the buffer.
	     
Our proposed system introduces the traffic profiling module (\figurename{ \ref{fig:System}}) that performs the following tasks in real time:
\begin{enumerate}
	\item \textit{Detection of the streaming flow:} A streaming flow is distinguished from other packet flows by its characteristic pattern of packet sizes and IAT. Once the system detects a filling phase followed by a steady-state phase (cp. \figurename{ \ref{fig:StreamingExample}}) a streaming flow is identified and referred to by its source IP address and destination TCP ports.

	\item \textit{Identification of the current streaming phase:} Filling and steady-state phases are identified by using combined criteria based on (i) streaming rate and (ii) characteristic burst patterns. These criteria are detailed in the next section.
		
	\item \textit{Extraction of parameters per streaming phase:} For each identified phase, basic parameters such as data volume, start time, end time, duration and rate can be extracted from the packet flows. Based on the rate match between streaming client and server in the steady-state phase, the observed streaming rate can be used as an estimator for video encoding rate.
\end{enumerate}
	
We assume that after the video starts playing, the user does not interrupt the video play-back by pausing or skipping forwards or backwards. These events can still be detected but are neglected for simplicity. 

\section{Video phase identification and rate estimation}
\label{sec:VRSE}
\begin{figure}[t]
	\centering
	\includegraphics[width=0.8\linewidth,trim={0cm 0cm 15.5cm 7cm},clip]{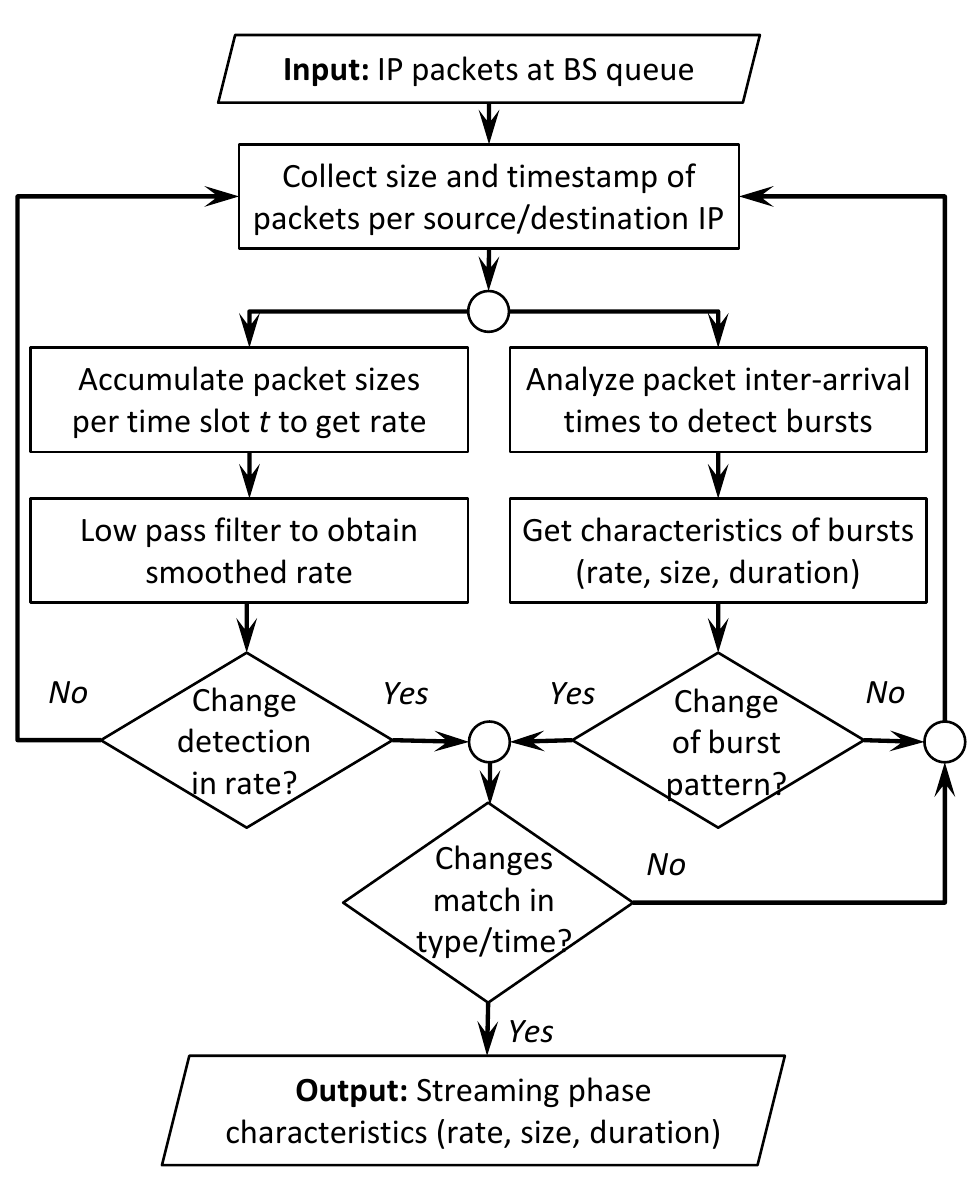} 
	\caption{Main decision flow of the traffic profiling module.}
	\label{fig:VRSE}
\end{figure}

In this section we specify the traffic profiling algorithm for the identification of streaming phases and the estimation of video encoding rate. The decision flow of the traffic profiling module is illustrated in Fig. \ref{fig:VRSE}. Firstly, from the downlink transmit queue, the size and the arrival time of the IP packets are observed per tuple of source and destination IP address. A video stream is identified by an initial filling and a subsequent steady-state phase. This requires to clearly distinguish these two phases from each other. To do so, we apply two methods in parallel and combine their results for the final decision. 

The first method is based on the streaming rate and aims to detect significant changes in data rate. For any arriving packet of index $k$, payload size $S_k$ is measured and aggregated within a time interval of duration $\Delta t$. For any interval of index $t$, this aggregation provides the streaming rate 
\begin{equation}
\rho_t=\frac{\sum_{k\in \mathcal{K}_t}S_k}{\Delta t},
\label{ratePerSlot}
\end{equation}
where $\mathcal{K}_t$ corresponds to the set of packets that arrived in this interval. To avoid misdetections due to small variation, we perform change detection on a smoothed time series. In our implementation, we obtain the smooth data rate $r^s_t$ from the regressive low-pass filter 
\begin{equation}
r^s_t=(1-a)r^s_{t-1}+a \rho_t,
\label{afilter}
\end{equation}
with attenuation factor $a$. We chose this filter due to its low complexity, but other filters or moving averages can be used as well.
Based on this smooth time series, significant rate changes can be detected with the help of quickest change detection algorithms \cite{Veeravalli} or by using simple threshold-based approaches, such as
	\begin{equation}
	f_t = \left\{\begin{array}{rl}
			1 & \mbox{if } r^s_t > c r^s_{t,max} \mbox{ and } f_{t-1} = -1 \\
			-1 & \mbox{if } r^s_t < \left(1-c\right)r^s_{t,max} \mbox{ and } f_{t-1} = 1 \\
			f_{t-1} & \mbox{otherwise,}
		\end{array} 
		\right.
		\label{method1} 
	\end{equation}
that we used in our studies, where $r^s_{t,max} = \max_{\tau \in \left[1,t\right]} r^s_\tau$ is the maximum smoothed rate until time slot $t$, factor $c$ is a constant and $f_t$ is a flag, with $f_0=-1$, that indicates rate changes, i.e., a rate increase at $t$ in case of a transition from $f_{t-1}=-1$ to $f_t=1$ and a rate decrease for an inverse transition.

The second method is based on detecting the steady-state phase, i.e., the on-off transmission pattern. For this reason, the packet IATs are analyzed to separate data bursts $b_n$, allowing per-burst characteristics, such as size $b^s_n$, duration $b^d_n$ and rate $b^r_n$, to be calculated. By analyzing this data, bursts can be classified to potential parts of a filling or a steady-state phase. For instance, long bursts and shorter repetitive bursts of similar duration are good indicators of a filling and a steady-state phase, respectively. Our studies showed that a static list of rules is more accurate and faster for burst classification than Support Vector Machines (SVMs). Consequently, we define bursts by a threshold $h_t$ for packet IAT and classify them by    
	\begin{equation}
	b_n = \left\{\begin{array}{rl}
	1 & \mbox{if } b^d_n \geq h_d \mbox{ and } b^r_n \geq h_r b^r_1 \\
	-1 & \mbox{if } b^d_n < h_d \mbox{ and } b^r_n \geq h_r b^r_1  \\
	0 & \mbox{otherwise,}
	\end{array} 
	\right.
	\label{method2} 
	\end{equation}
where $h_d$ and $h_r$ are thresholds for burst duration and rate, and $b_n$ is a flag that indicates a potential burst of a filling and a steady-state phase for $b_n=1$ and $b_n=-1$ respectively. Note that $b^r_1$ is used here to indicate the rate of the initial filling phase. Bursts of few packets, that could otherwise affect the output of the algorithm, are discarded by setting a threshold  $h_s$ of few KB for the size of the burst and moreover, $h_n$ consecutive bursts with $b_n=-1$ must be found to identify a steady-state phase.   

As given in Fig. \ref{fig:VRSE}, the output of both methods is compared in terms of (i) type of state (filling, steady-state) and (ii) estimated time of change, by allowing a short deviation in the order of a few seconds. 
If the results match, the algorithm provides the characteristics of a streaming phase, such as the average video bit-rate per streaming phase and the size of downloaded video data. For any new input, the process is repeated in order to detect new streaming phases. 

The combination of the above two methods improves the accuracy of the detection. For example, if the second method is not applied and the throughput drops in the middle of a filling phase, the last part of this phase may be mistakenly interpreted as a steady-state phase. 

\section{Experimental validation}
\label{sec:results}
In this section, we present the setup, methodology and results of our experimental validation. The accuracy of our traffic profiling solution was studied for the popular streaming service YouTube on two mobile devices and compared with the ground truth, i.e. the actual streaming phase and video encoding rate. The performance comparison to DPI is left for our future work. Overall, we performed 960 experiments over the course of one week, representing more than 170 hours of studied streaming traffic. 

\subsection{Setup}
\label{sec:setup}
\begin{figure}[t]
	\centering
	\includegraphics[width=.95\linewidth,trim={12cm 0cm 0cm 0cm},clip]{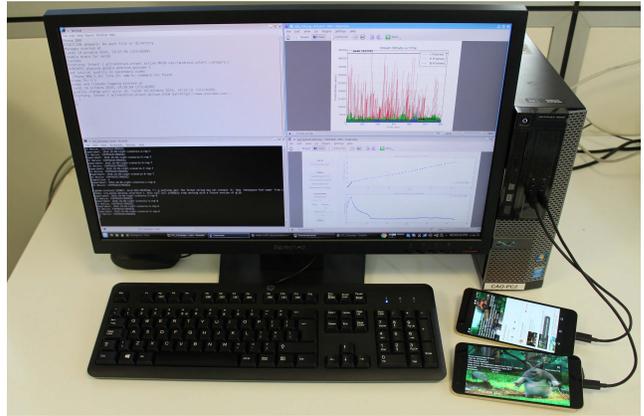} 
	\caption{Experimental setup with two Smartphones connected to the Internet through a computer acting as WLAN access point. The Smartphones are also connected via USB cables for controlling the experiments.}
	\label{fig:ExpSetupPhoto}  
\end{figure}

In order to measure YouTube traffic in an automatic and reproducible manner, we designed the testbed in \figurename{ \ref{fig:ExpSetupPhoto}}. Two Android Smartphones (Huawei Nexus 6P, baseband version: angler-03.61, Android 6.0.1 with the security patch from June 1\textsuperscript{st}, 2016) are connected via a Wireless Local Area Network (WLAN) to a Linux computer (Kernel 3.16.0-71-lowlatency) that operates as a WLAN access point. The computer is connected to the Internet via a T1 line, acts as a gateway for the Smartphones, and controls the phones via a Universal Serial Bus (USB) connection. The WLAN operates in IEEE 802.11g mode at a carrier frequency of 2412 MHz. Due to the close distance between phones and access point (cp. \figurename{ \ref{fig:ExpSetupPhoto}}), the Signal-to-Interference-plus-Noise Ratio (SINR) was 23 dB, which provides the maximum physical layer rate of 54 Mbit/s.

Selecting WLAN instead of Long-Term Evolution (LTE) allowed us to directly configure rate throttling in a reproducible manner. To do so, we used the traffic configuration (tc) tool provided in the Linux kernel and controlled it by shell scripts. Layer 3 and Layer 4 packet logs were then recorded with tcpdump \cite{tcpdump}, although Layer 4 information is not used by our traffic profiling system. The traffic was generated with the  native YouTube application (version: 10.28.60) for Android. According to \cite{Wamser2016} and our observations, this version performs standard DASH operation \cite{DASH}. The YouTube application protects its streaming traffic via TLS and consequently, HTTP queries are sent to the server TCP port 443. Over the course of our measurement campaign, the average round-trip-time (RTT) was 28 ms. Note that YouTube's front-end occasionally changed the streaming servers, sometimes even for the same video. These changes, however, were made within a small set of servers and had no significant effect on RTT and throughput. To this end, we decided to study the streams of all servers in a single data set.

To validate the estimated video bit-rate, the true video encoding rate was obtained by downloading the video segments of the streamed DASH representation from the responding YouTube server and by decoding it with the ffprobe statistics tool \cite{ffmpeg}. As streaming content, we used the open movies Big Buck Bunny (BBB) \cite{BigBuckBunnieMovie} and Tears of Steel (TOS) \cite{TearsOfSteelMovie}. Both movies are commonly used for testing video codecs and streaming protocols and recommended in the measurement guidelines of the DASH Industry Forum \cite{Forum2014}. BBB is of 9:56 min duration, high motion, encoded at 24 fps with the H.264 codec in an MP4 container. We used the DASH representations 480p and 720p, with an average encoding rate of 646 Kbps and 1346 Kbps, respectively. For TOS, we used the same representations with average encoding rates of 686 Kbps and 1253 Kbps. This movie has a duration of 12:14 min and is encoded in the same format as BBB. The average audio bitrate is 128 Kbps for both movies.

\subsection{Scenarios} 
For each of the two movies, we tested 4 different scenarios per video with 120 repetitions per scenario, which were conducted in Paris, France, over one week in October 2016. To assure full batteries, both phones measured in alternating cycles of 12 hours. The 4 studied scenarios are specified in Table \ref{table:scenarios} and include the following cases: 
\begin{enumerate}
	\item \textit{MQ}: Medium Quality (480p) for the entire video.
	\item \textit{HQ}: High Quality (720p) for the entire video.
	\item \textit{QC}: Quality Change (720p to 480p) at a random time in the interval [120, 240] s of the video. 
	\item \textit{AQ}: Adaptive Quality selection with 90 s of rate throttling to 320 Kbps, starting at a random time in the interval [120, 240] s.
\end{enumerate}

Scenarios MQ and HQ are chosen in order to study the performance of the algorithm in simple cases with a single filling and steady-state phase, for two quality levels with significant bit-rate difference. Then, QC is selected to verify that the algorithm can also detect multiple transitions between streaming phases throughout the video session. Finally, AQ is a more challenging scenario due to the introduced buffer depleting phases as a result of limited throughput. In this case, YouTube's HAS policy automatically chooses the video quality throughout the session.

\begin{table}[!t]
	\renewcommand{\arraystretch}{1.3}
	\caption{Experimental Scenarios}
	\label{table:scenarios}
	\centering
	\begin{tabular}{|c|c|c|c|}
		\hline
		\bfseries Scenario & \bfseries Initial & \bfseries Quality & \bfseries Throughput \\
		\bfseries name & \bfseries quality & \bfseries change & \bfseries limitation \\  
		\hline\hline
		MQ & 480p & No & No   \\ 
		\hline
		HQ & 720p & No & No   \\ 
		\hline
		QC & 720p & Yes (720p$\rightarrow$480p) & No \\ 
		\hline
		AQ & Auto & Auto & Yes (320 kbps) \\ 
		\hline
	\end{tabular}
\end{table}

\subsection{Results}
We begin with some general observations from the experimental results. First, we noticed that during steady-state phases, the ratio of streaming to encoding rate, known as \emph{throttling factor}, is approximately equal to 1. This differs from the results in \cite{Ameigeiras2012,CharMobileYouTube,Rao2011} that show a throttling factor equal to 1.25 and even higher in some cases. Furthermore, we noticed that the YouTube streaming algorithm aims to keep the buffer level constant around a fixed level of 18 MB for all the studied videos and qualities, as shown in \figurename{ \ref{fig:StreamingExample}}. This is different from the findings in the recent work in \cite{Wamser2016}, where the buffer is stabilized around a fixed value of 50 s. It is also worth mentioning that in 98\% of the experiments, a single server provided the video stream over a session, while two streaming servers were used in the remaining 2\%.

The measurement results for the proposed video phase identification and rate estimation are presented below. The results were obtained with the free parameters (cp. Section \ref{sec:VRSE}) $\Delta t = 0.1$ s, $a=0.02$, $c=0.6$, $h_t=1.5$ s, $h_d=5$ s, $h_r=0.3$, $h_s=20$ KB and $h_n=3$, which was a robust choice according to our measurements.

\begin{figure}[t]
	\centering
	\includegraphics[width=1\linewidth,trim={.8cm 0cm 1.2cm 0.5cm},clip]{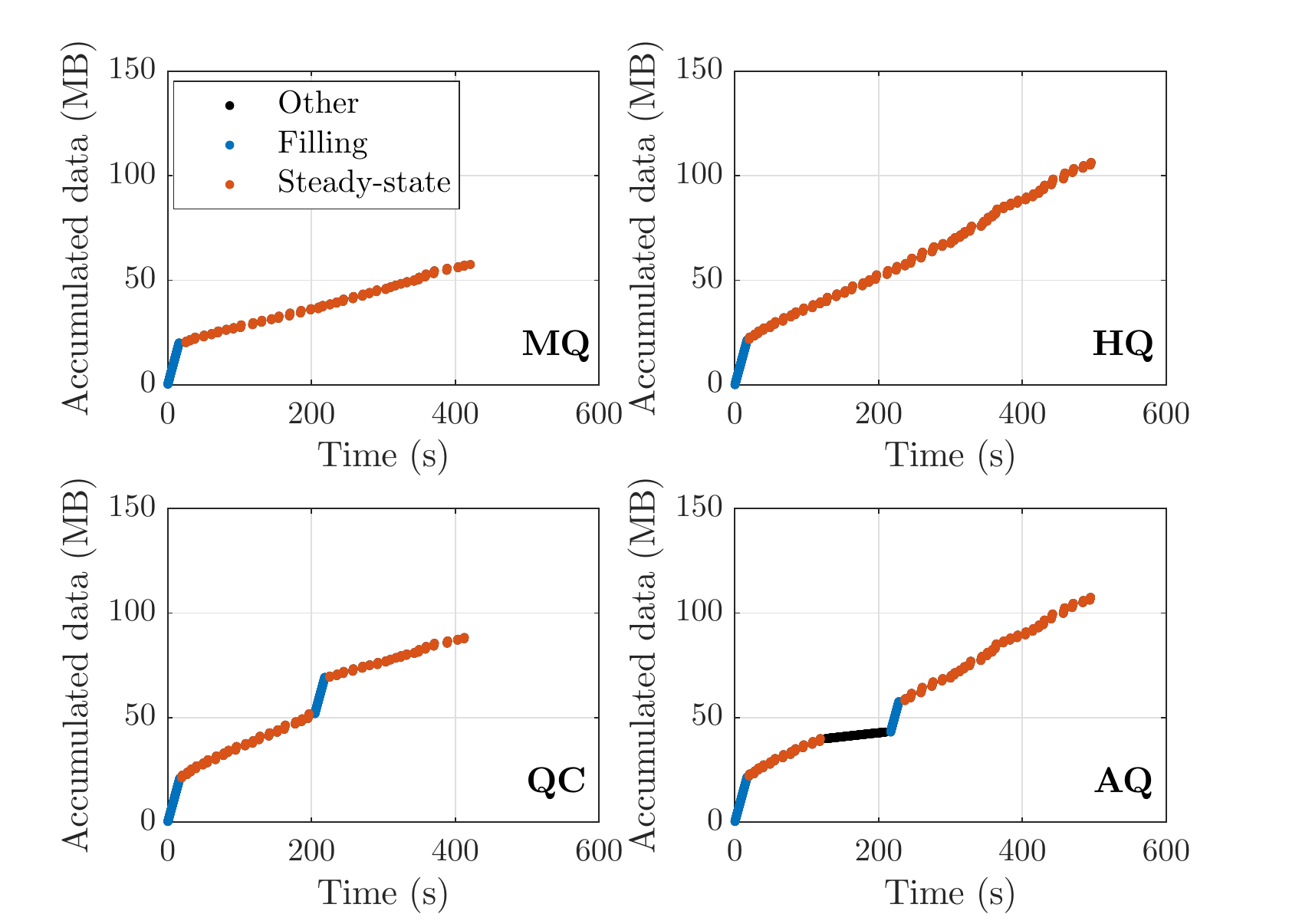} 
	\caption{Big Buck Bunny movie: examples of automatic phase identification for the 4 studied scenarios.}
	\label{fig:StateDetectionExamplesBBB}  
\end{figure}

\begin{figure}[t]
	\centering
	\includegraphics[width=1\linewidth,trim={.8cm 0cm 1.2cm 0.5cm},clip]{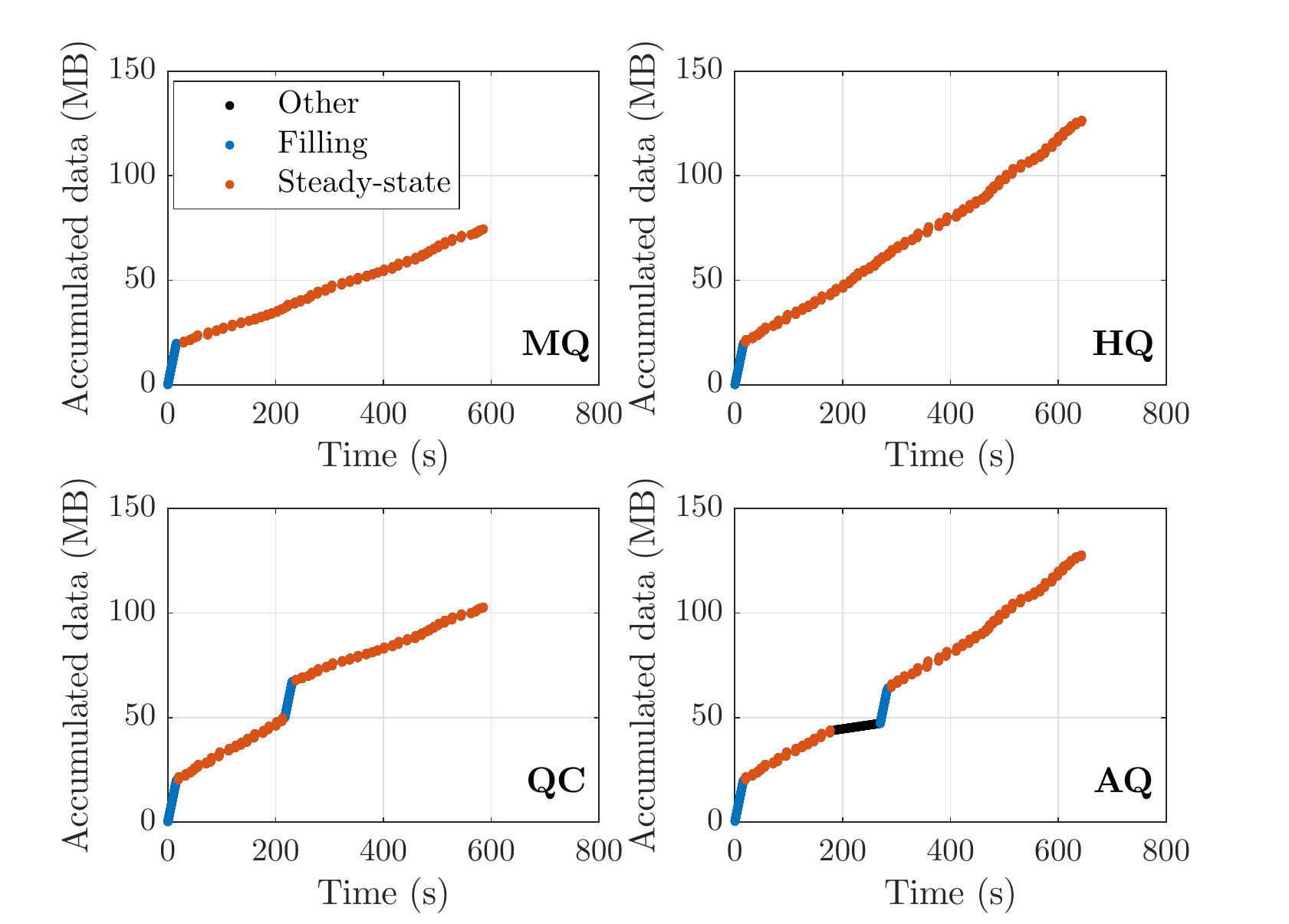} 
	\caption{Tears of Steel movie: examples of automatic phase identification for the 4 studied scenarios.}
	\label{fig:StateDetectionExamplesTOS}  
\end{figure}

\subsubsection{Phase identification}
\figurename{ \ref{fig:StateDetectionExamplesBBB}} and \figurename{ \ref{fig:StateDetectionExamplesTOS}} present examples of phase identification for the two studied movies, respectively. In these figures, traffic profiling already classifies the streaming phase automatically. We have also introduced a third category, denoted as \emph{other}, in order to distinguish remaining parts of streaming that do not belong in the \emph{filling} or \emph{steady-state} categories. 
For both movies, similar behavior is observed apart from small differences in total video duration and size. Regarding MQ and HQ scenarios, we can easily notice the increased slope in HQ due to higher video bit-rate. More interestingly, in the QC scenario, a second filling phase is detected after the manual quality change, followed by a new steady-state phase. This is expected as the application tries to fill the buffer with the new quality. The decreased slope in the second steady-state also verifies that a lower quality is selected. Finally, in the AQ scenario, a second filling phase is triggered in order to refill the buffer up to the target level, after the period of limited throughput. This exact period is classified as \emph{other}, since $b_n=0$ according to \eqref{method2}, due to the rate criterion therein.         

A confusion matrix that summarizes the performance of the phase identification over the complete set of experiments is presented in Table \ref{table:ConfusionMatrix2}. The proposed algorithm achieves a very high accuracy for the studied scenarios by correctly detecting each phase with a percentage greater than 99\%. The few mis-classifications were due to (i) multiple source IP addresses or (ii) strong throughput dynamics due to network congestion. We leave the further improvement of the algorithm for these specific cases for our future work. 

\begin{table}[!t]
	\renewcommand{\arraystretch}{1.3}
	\caption{Confusion Matrix}
	\label{table:ConfusionMatrix2}
	\centering
	\begin{tabular}{cc|c|c|c|}
		\cline{3-5}
		& & \multicolumn{3}{ c| }{\bfseries Identified} \\ \cline{2-5}
		& \multicolumn{1}{ |c|  }{(\%)} & filling & steady-state &  other \\ \cline{1-5}
		\multicolumn{1}{ |c  }{\multirow{3}{*}{\rotatebox[origin=c]{90}{\bfseries True}} } &
		\multicolumn{1}{ |c| }{filling} & 99.2 & 0.1 & 0.7 \\ \cline{2-5}
		\multicolumn{1}{ |c  }{}                        &
		\multicolumn{1}{ |c| }{steady-state} & 0.0 & 99.0 & 1.0 \\ \cline{2-5}
		\multicolumn{1}{ |c  }{}                        &
		\multicolumn{1}{ |c| }{other} & 0.1 & 0.0 & 99.9 \\ \cline{1-5}
	\end{tabular}
\end{table}
		
\subsubsection{Rate estimation}
Having confirmed the high accuracy of phase identification, we proceed with studying the performance of rate estimation. To this end, we compare the estimated video rate $\hat r$, calculated as the average streaming rate during the identified steady-state phase, against the average true video bit-rate $r$ during this steady-state. Since the time interval of this phase is not exactly the same over all experiments, $r$ is not a constant value. 

\begin{figure}[t]
	\centering
	\includegraphics[width=.95\linewidth,trim={.8cm 0cm 1.2cm 0.5cm},clip]{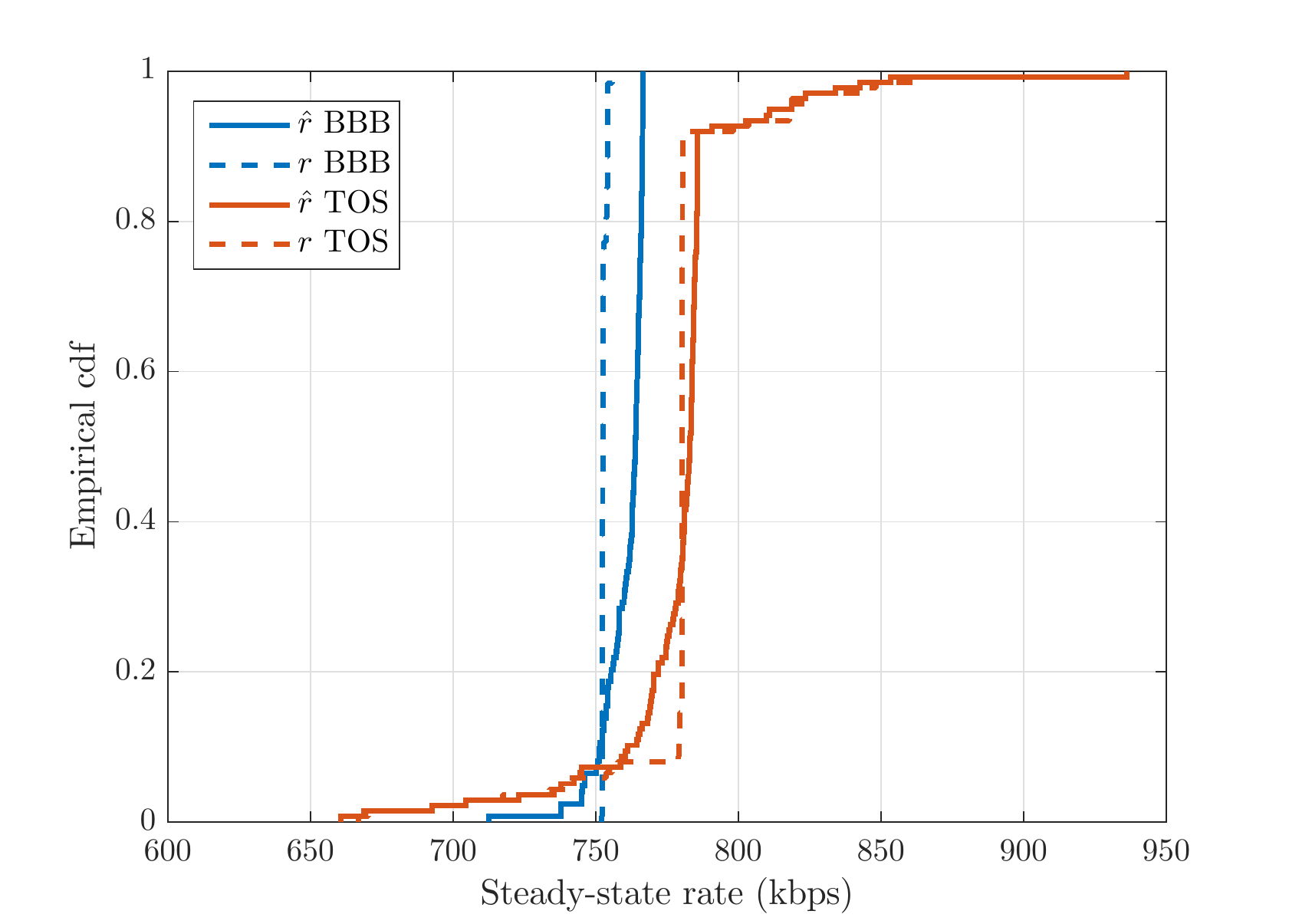} 
	\caption{MQ scenario: empirical cdf for both BBB and TOS videos; solid lines for estimated bit-rate and dashed lines for true video bit-rate.}
	\label{fig:ecdfMQ}
\end{figure}

\begin{figure}[t]
	\centering
	\includegraphics[width=.95\linewidth,trim={.8cm 0cm 1.2cm 0.5cm},clip]{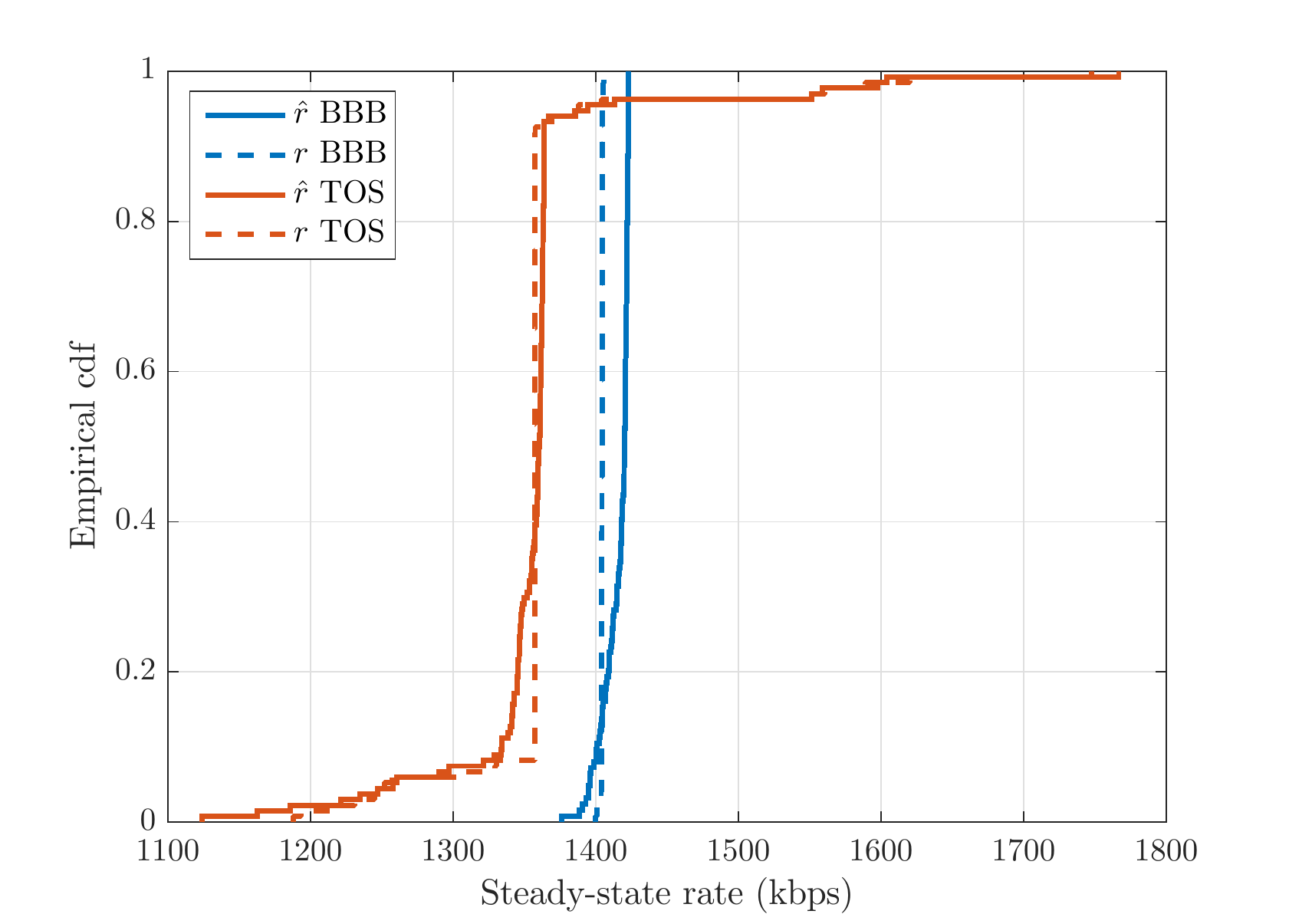} 
	\caption{HQ scenario: empirical cdf for both BBB and TOS videos; solid lines for estimated bit-rate and dashed lines for true video bit-rate}
	\label{fig:ecdfHQ}  
\end{figure}

\begin{figure}[t]
	\centering
	\includegraphics[width=.95\linewidth,trim={.8cm 0cm 1.2cm 0.5cm},clip]{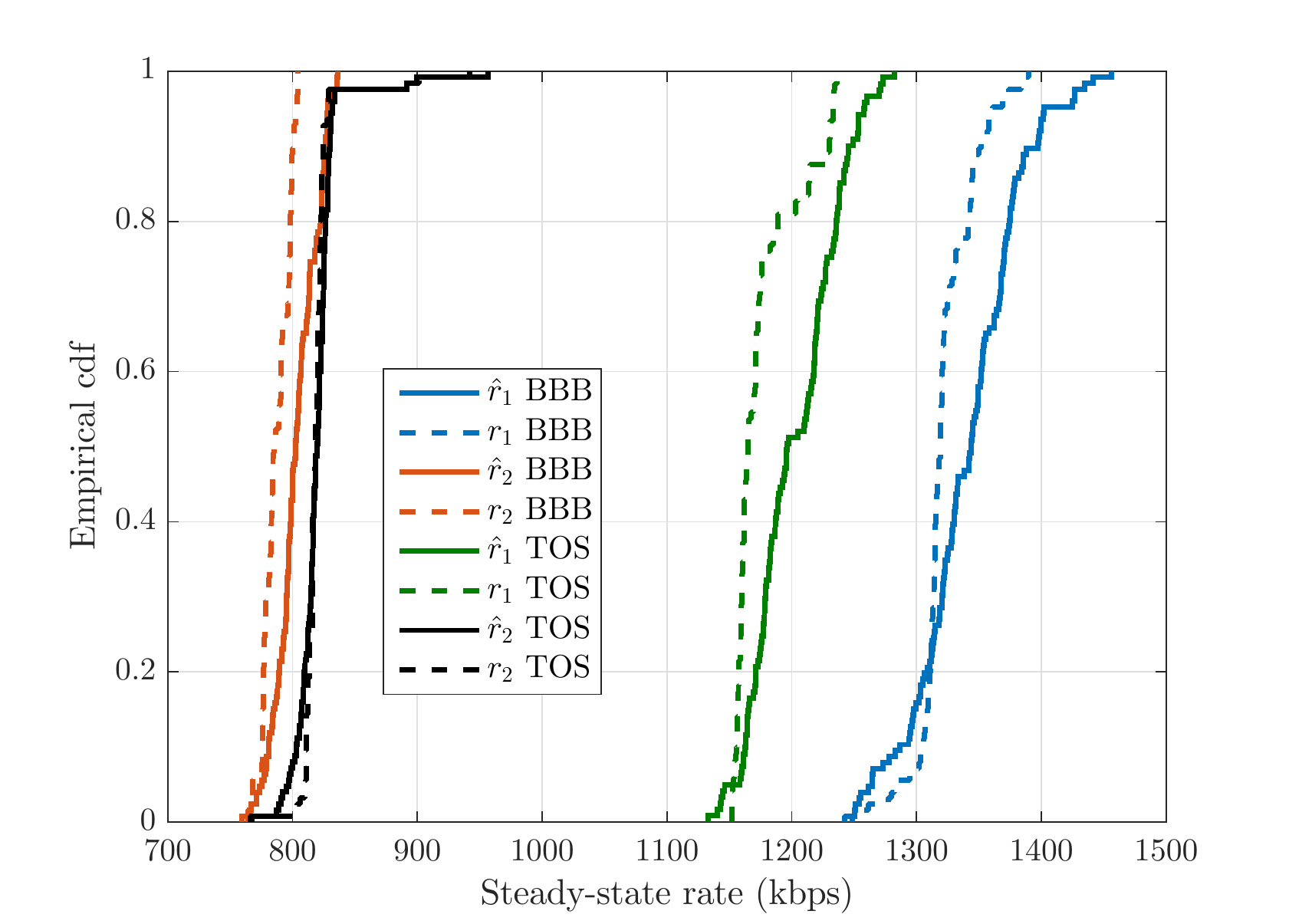} 
	\caption{QC scenario: empirical cdf for both BBB and TOS videos; results shown for both steady-state phases per video.} 
	\label{fig:ecdfQC}  
\end{figure}

\begin{figure}[t]
	\centering
	\includegraphics[width=.95\linewidth,trim={.8cm 0cm 1.2cm 0.5cm},clip]{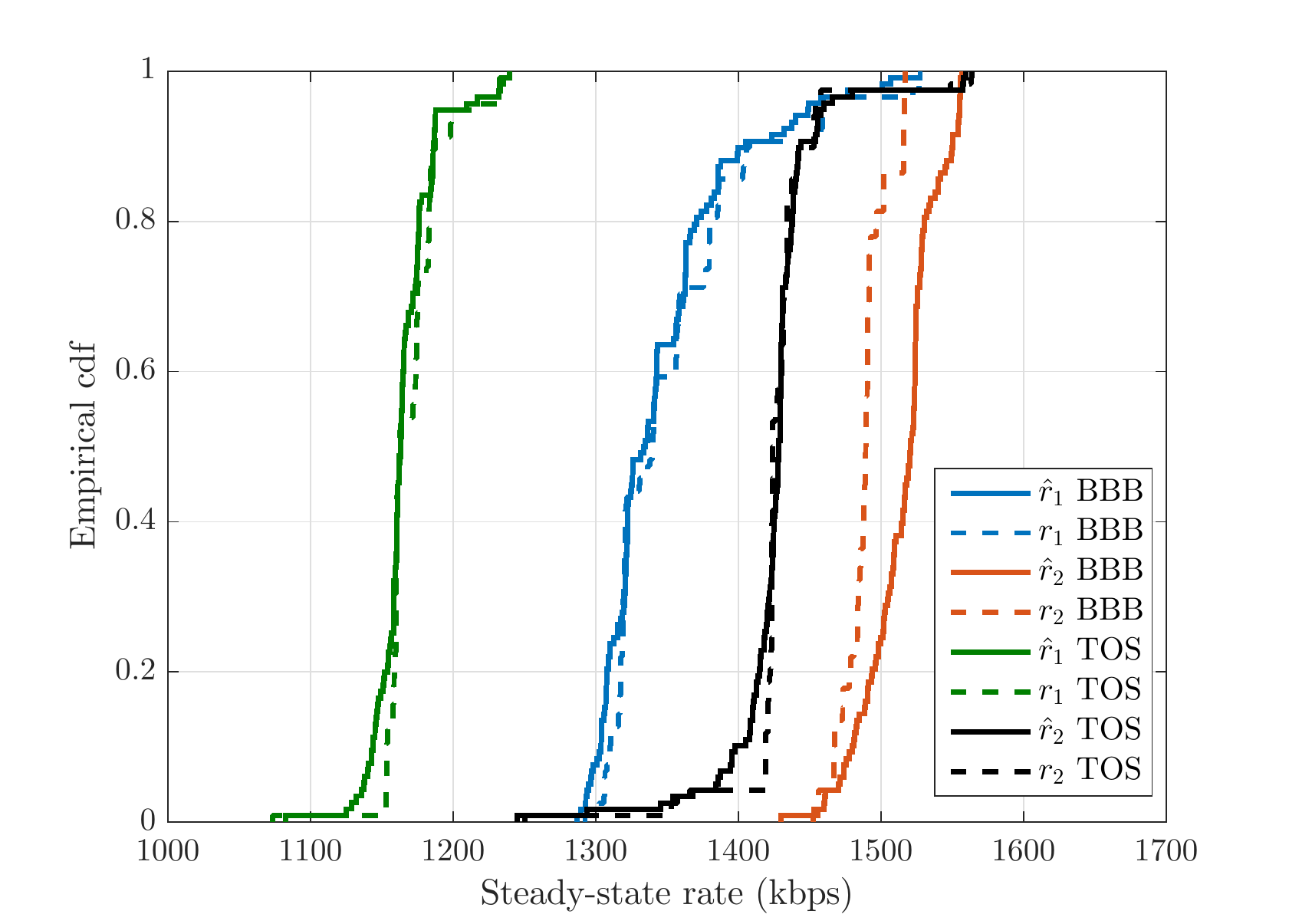} 
	\caption{AQ scenario: empirical cdf for both BBB and TOS videos;  results shown for both steady-state phases per video.}
	\label{fig:ecdfAQ}  
\end{figure}

\figurename{ \ref{fig:ecdfMQ}} and \figurename{ \ref{fig:ecdfHQ}} present the empirical cumulative distribution function (cdf) for the MQ and HQ scenarios, respectively. It is clear that the estimated rate is very close to the true video bit-rate, with a slight over-estimation in the case of the BBB movie. Moreover, we can see that the cdfs for TOS span over a larger range of values, indicating the higher dynamic of this movie. The next numerical results for QC and AQ scenarios concern multiple steady-state phases, as already shown in \figurename{ \ref{fig:StateDetectionExamplesBBB}} and \figurename{ \ref{fig:StateDetectionExamplesTOS}}. Let us define by $\hat r_1$, $r_1$ the estimated and true video bit-rate respectively for the first steady-state phase, and similarly by $\hat r_2$, $r_2$ those for the second phase. \figurename{ \ref{fig:ecdfQC}} presents the empirical cdfs for both videos in the QC scenario, where we can also verify the high accuracy of the estimation. Similar conclusions apply to \figurename{ \ref{fig:ecdfAQ}} for the AQ scenario, where it is interesting to observe the increased rate during the second steady-state phase, due to higher motion scenes in the last minutes of both videos.


\begin{figure}[t]
	\centering
	\includegraphics[width=.95\linewidth,trim={.8cm 0.1cm 1.2cm 0.4cm},clip]{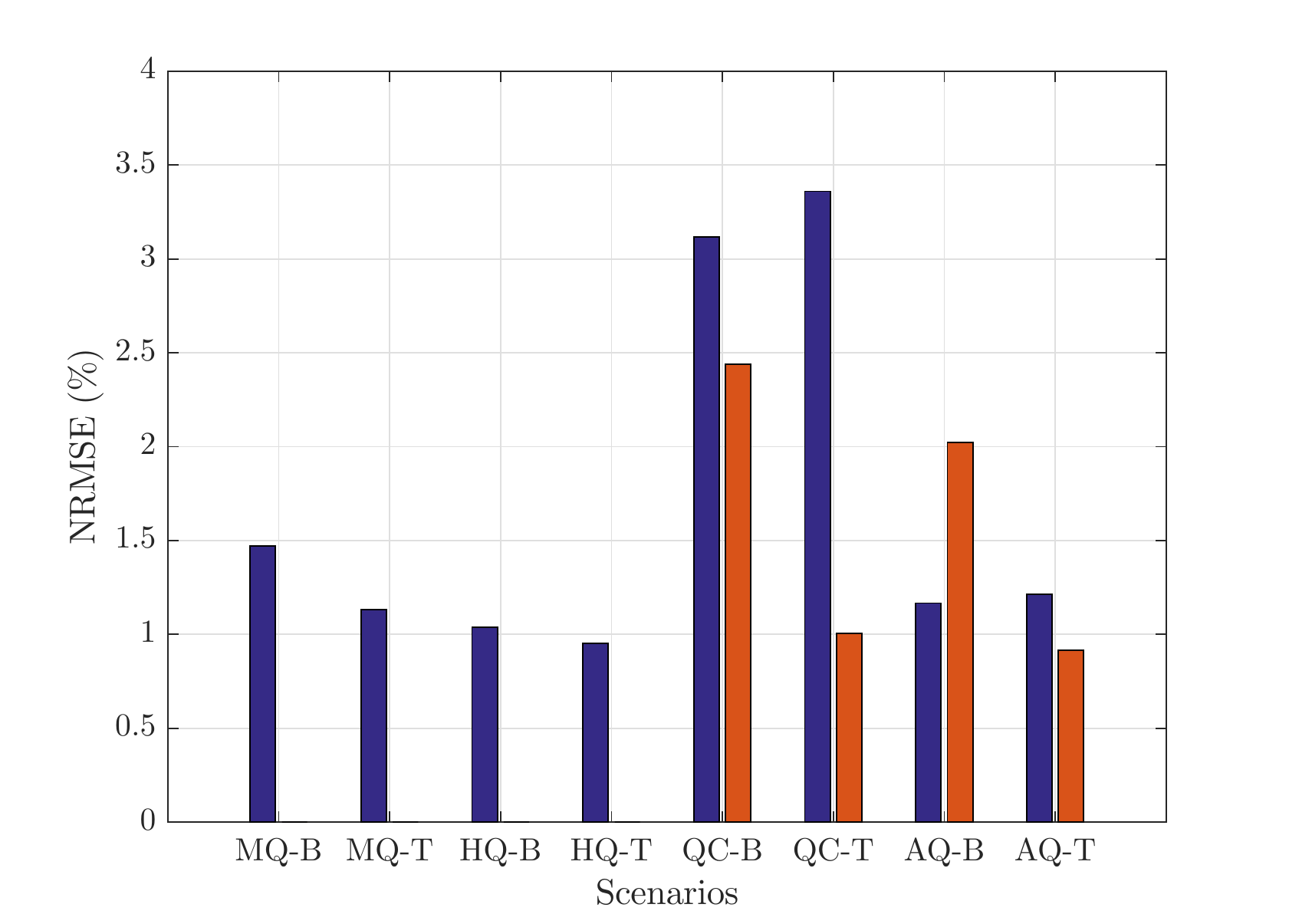} 
	\caption{NRMSE for all studied scenarios; results per steady-state phase are presented in the scenarios with multiple steady-state phases. }
	\label{fig:NRMSE}  
\end{figure}

Finally, \figurename{ \ref{fig:NRMSE}} summarizes the performance of rate estimation for all scenarios in terms of normalized root-mean-square error (NRMSE). We use suffixes '-B' and '-T' to indicate the BBB and TOS movie, respectively, and normalize by the average $r$ over all experiments of the corresponding scenario. An impressive accuracy of less than 2\% NRMSE for MQ, HQ and even for the more challenging AQ scenario is shown. The accuracy drops slightly for the QC scenario, where the NRMSE reaches values of 3.5\% during the first steady-state phase. This results from the fact that the manual quality change forces new bursts of data to be transmitted earlier than expected, leading to a marginal over-estimation of the rate. This bias can be easily removed by discarding the last data burst of the steady-state from the rate calculations. 

As a final remark, the reader should keep in mind that all the rate estimation results can be further improved if we consider the start-up delay, which leads to a small offset between the time intervals of $\hat r$ and $r$.

 
\section{Conclusions}
We introduced traffic profiling to extract application-layer characteristics from on-going video streams at the network or link layer. By observing the packet flows at either of these layers, our profiling system detects the characteristic phases of an HTTP adaptive streaming session in order to estimate the state of the play-back buffer and the video encoding rate. Our extensive experimental validation includes more than 170 hours of YouTube traffic and shows very high accuracy for the proposed state and rate estimation method. As traffic profiling works even with TLS/SSL encryption and does not rely on standardized signaling interfaces, it provides a valuable alternative to DPI and cross-layer signaling.

Our traffic profiling system can be readily applied to provide video information to network optimization functions such as traffic-aware schedulers, admission control and dynamic routing. Moreover, QoE reporting and traffic shaping mechanisms, as recently deployed for video traffic \cite{bingeonPaper}, may highly profit from the accurate estimation.

As future work, we will apply Quickest Change Detection methods \cite{Veeravalli} for faster detection, investigate traffic profiling for the Quick UDP Internet Connections (QUIC) protocol \cite{quic}, and compare the performance of our solution to DPI.

\label{sec:conclusions}

\bibliography{IEEEabrv,VRSE}
\bibliographystyle{IEEEtran}

\end{document}